# Rapidity dependent transverse momentum spectra of heavy quarkonia produced in small collision systems at the LHC

Li-Na Gao[1;1)], Fu-Hu Liu[2;2)], Bao-Chun Li[2;3)]

[1] *Department of Physics, Taiyuan Normal University, Jinzhong, Shanxi 030619, China*

[2] *Institute of Theoretical Physics and State Key Laboratory of Quantum Optics and Quantum Optics Devices, Shanxi University, Taiyuan, Shanxi 030006, China*

**Abstract:** The rapidity dependent transverse momentum spectra of heavy quarkonia ($J/\psi$ and $\Upsilon$ mesons) produced in small collision systems such as proton-proton ($pp$) and proton-lead ($p$-Pb) collisions at center-of-mass energy (per nucleon pair) $\sqrt{s}$ ($\sqrt{s_{NN}}$) = 5-13 TeV are described by a two-component statistical model which is based on the Tsallis statistics and inverse power-law. The experimental data measured by the LHCb Collaboration at the Large Hadron Collider (LHC) are well fitted by the model results. The related parameters are obtained and the dependences of parameters on rapidity are analyzed.

## 1. Introduction

The study of high energy proton-proton, proton-nucleus and nucleus-nucleus collisions [1-5] can provide a unique opportunity for ones to understand the strong interaction theory and nuclear reaction mechanism [6-10], and analyze the evolution processes of interacting system and quark-gluon plasma (QGP). At the same time, by this study, one can examine the standard model and other phenomenological models or statistical methods [11-14], and search for new physics beyond the standard model. This study also provides new information for people to understand the origin of the universe. As the basic element in nuclear collisions, proton-proton collisions are worthy to study. Meanwhile, as a transition from proton-proton collisions to nucleus-nucleus collisions, proton-nucleus collisions are also worthy to study.

With the development of modern experimental and detecting technology, the collision energy has been continuously improved. Meanwhile, more and more information about collision process can be accurately measured in experiments [15-19]. Because the collision time of interacting system is very short, one can only analyze the characteristics of final particles produced in the collisions to obtain the mechanisms of nuclear reactions and the properties of formed matter such as QGP.

Generally, the information of nuclear reactions in experiments can be obtained by measuring the transverse momentum spectrum and correlation, pseudorapidity or rapidity spectrum and correlation, anisotropic flow distribution and correlation, multiplicity distribution and correlation, nuclear modified factor, and so forth [15-19]. The transverse momentum spectrum is one of the most general objects in the study. It is measured by experiments and provides information about temperature and excitation degree of interacting system at the stage of kinetic freeze-out. Therefore, the study of transverse momentum spectrum of final particles is greatly significative

---

[1)] E-mail: gao-lina@qq.com; gaoln@tynu.edu.cn
[2)] E-mail: fuhuliu@163.com; fuhuliu@sxu.edu.cn
[3)] E-mail: libc2010@163.com; s6109@sxu.edu.cn

in analyzing the mechanisms of nuclear reactions and the properties of QGP.

Many theoretical models and formulas have been applied for the descriptions of transverse momentum spectra. These models and formulas include, but are not limited to, the Boltzmann-Gibbs statistics [1-3], Lévy distribution [4, 5], Erlang distribution [6], Tsallis statistics [7-14], and so on. In this paper, we use a two-component statistical model to describe the experimental transverse momentum spectra of heavy quarkonia ($J/\psi$ and $\Upsilon$ mesons) produced in small collision systems such as proton-proton ($pp$) and proton-lead ($p$-Pb) collisions. The data quoted by us are measured by the LHCb Collaboration [15-18] at the Large Hadron Collider (LHC), though other data are available [19]. The two-component statistical model is based on the Tsallis statistics and inverse power-law.

In the following sections, we describe the formulism of the two-component statistical model in section 2. The results and discussion are given in section 3. Finally, the conclusions of the present work are given in section 4.

## 2. The formulism

Within the framework of the multi-source thermal model [20-22], the emission sources of final particles produced in high energy collisions can be divided into several groups due to different interacting mechanisms, impact parameter ranges (centrality classes), or event samples. A typical classification is soft excitation and hard scattering processes [23-26], and even including very-soft excitation and very-hard scattering processes. Generally, one can use different models and formulas to describe different processes. In some cases, one can use the same model and formula to describe different processes. In other cases, one can use different models and formulas to describe the same process.

The Tsallis statistics has been widely applied for high energy collisions [27-31]. It describes different particle spectra in different processes, but not the heavy quarkonium spectra in very-hard process in some cases. For the soft and very-soft processes, the Boltzmann-Gibbs statistics [1-3] also play a main role in the description. For the hard and very-hard processes, an inverse power-law [32-35] play the main role in the description. For the transverse momentum ($p_T$) spectra of heavy quarkonia ($J/\psi$ and $\Upsilon$ mesons) produced in collisions at the LHC, we need a superposition of the Tsallis statistics and the inverse power-law, which is a two-component statistical model.

In the Tsallis statistics [27-30], the invariant momentum ($p$) distribution is

$$E\frac{d^3N}{dp^3} = \frac{gV}{(2\pi)^3} m_T \cosh y \left[1 + \frac{q-1}{T}(m_T \cosh y - \mu)\right]^{-q/(q-1)}, \qquad (1)$$

where $E$ is the energy, $N$ is the particle number, $g$ is the degeneracy factor, $V$ is the volume, $m_T = \sqrt{p_T^2 + m_0^2}$ is the transverse mass, $m_0$ is the rest mass, $T$ is the temperature parameter, $q$ is the entropy index, and $\mu$ is the chemical potential. The normalized $p_T$ distribution can be given by

$$\frac{1}{N}\frac{dN}{dp_T} = \frac{gV}{(2\pi)^2} p_T m_T \int_{y_{\min}}^{y_{\max}} \cosh y \left[1 + \frac{q-1}{T}(m_T \cosh y - \mu)\right]^{-q/(q-1)} dy. \qquad (2)$$

In the mid-rapidity ($y = 0$) region, the formulism of Tsallis statistics can be given by

[31]

$$f_1(p_T) = C_T p_T m_T \left[1 + \frac{q-1}{T}(m_T - \mu)\right]^{-q/(q-1)}, \tag{3}$$

where $C_T = gV/(2\pi)^2$ is the normalization constant related to the free parameters. When the collision energy is high enough, the chemical potential is especially small. In the energy range of LHC, the value of $\mu$ approximately is zero [27-29].

In some cases, the experimental data are presented in a given rapidity range, which is generally not in the mid-rapidity region. We have to shift simply the given rapidity range to the mid-rapidity region by subtracting the mid-value of the given rapidity range, and use Eq. (3) directly. If we consider the differences of rapidities in the given rapidity range or in the mid-rapidity region, a more accurate Eq. (2) which includes the integral for the rapidity can be used. If we consider the given rapidity range in the more accurate Eq. (2), the kinetic energy of directional movement will be included in the temperature, which causes a larger temperature and is not correct. In fact, in the mid-rapidity region, the difference between the minimum (maximum) rapidity and 0 is neglected. The more accurate Eq. (2) is not needed.

It should be noted that when we use the multi-source thermal model and the Tsallis statistics, each group or process is assumed to stay in a local equilibrium state. The excitation degree of each group or process is described by the temperature parameter $T$, and the equilibrium degree is described by the entropy index $q$. A large $T$ corresponds to a high excitation degree, and a large $q$ ($q \gg 1$) corresponds to a far away from the equilibrium state. The closer to 1 the $q$ is, the closer to equilibrium the group or process becomes. In an equilibrium state, one has $q = 1$. Generally, $q$ is not too large. This means that each group or process stays approximately in a local equilibrium state.

The inverse power-law can describe the hard and very-hard processes. In refs. [32-34], the inverse power-law is described by the Hagedorn function [35], its parameterized form is expressed as

$$f_2(p_T) = A p_T \left(1 + \frac{p_T}{p_0}\right)^{-n}, \tag{4}$$

where $p_0$ and $n$ are free parameters, and $A$ is the normalization constant related to the free parameters.

In the Hagedorn function, scattering between nucleons may be thought of in terms of valence quarks. To measure the scattering strength, the parameters $p_0$ and $n$ can be used. A large $p_0$ and a small $n$ describe a wide $p_T$ range which means a violent scattering. Impact between quarks may also be described via pQCD (perturbative quantum chromodynamics), which gives an inverse power-law $p_T$ spectrum [32-34] which is the same as the Hagedorn function [35]. The pQCD also gives rapidity dependent $p_T$ spectra which results in rapidity dependent $p_0$ and $n$.

According to Eqs, (3) and (4), we can structure a superposition of the Tsallis statistics and the inverse power-law, which results in a two-component statistical model as

$$f(p_T) = k f_1(p_T) + (1-k) f_2(p_T), \tag{5}$$

where $k$ is the contribution ratio of the first component. Naturally, Eq. (5) is normalized to 1 due to the fact that Eqs. (3) and (4) are normalized to 1. Although the Tsallis statistics has more than one forms and the inverse power-law has different modified forms, we shall not discuss them further. In fact, Eq. (5) structured through Eqs. (3) and (4) is enough to use in the present work.

It should be noted that there are two types of superposition for two components. Except for Eq. (5), another superposition is the step function or the Hagedorn model [35]

$$f(p_T) = A_1 \theta(p_1 - p_T) f_1(p_T) + A_2 \theta(p_T - p_1) f_2(p_T), \tag{6}$$

where $A_1$ and $A_2$ are constants which ensure the contributions of two components are the same at $p_T = p_1$, and $\theta(x) = 1$ if $x > 0$ and $\theta(x) = 0$ if $x < 0$. Although there are entanglements in determining parameters by Eq. (5), the curve at $p_T = p_1$ is not smooth due to Eq. (6). Our very recent work [36] shows that Eqs. (5) and (6) result in similar values of parameters, especially for the trends. To obtain a smooth curve, Eq. (5) is used in the present work.

For a real fit process, we may select firstly a set of free parameters. Then, we may use the selected set of parameters in Eqs. (3) and (4), and let the two equations be normalized to 1 respectively. The normalization constants $C_T$ and $A$ can be determined and used back in Eqs. (3) and (4) so that the two equations can be used in Eq. (5). In the determination for the parameters, the method of least squares can be used. The errors of the parameters can be determined to let the confidence levels of fittings are 95% in most cases and 90% in a few cases if existent.

## 3. Results and discussion

Figure 1 shows the transverse momentum spectra, $d^2\sigma(J/\psi)dp_T dy$, of $J/\psi$ mesons produced in *pp* collisions at center-of-mass energy $\sqrt{s} = 7$ TeV, where $\sigma$ denotes the cross section. Figures 1(a)-1(d) present the results of the prompt $J/\psi$ with no polarisation, $J/\psi$ from *b* with no polarisation, prompt $J/\psi$ with full transverse polarisation, and prompt $J/\psi$ with full longitudinal polarisation, respectively. The symbols represent the experimental data measured by the LHCb Collaboration [15] at the LHC. In order to see clearly, different symbols are used to distinguish the different rapidity ranges in the panels. The curves are our results fitted by Eq. (5). The values of free parameters ($k$, $T$, $q$, $p_0$, and $n$) and $\chi^2$/dof (degree of freedom) corresponding to each curve in Figure 1 are listed in Table 1, where the normalization constants which reflect the areas under the curves are not listed to avoid trivialness. For the same reason, the concrete confidence levels are not listed in the table one by one. One can see that the experimental data measured by the LHCb Collaboration are well fitted by the two-component statistical model. The behaviors of parameters will be discussed later.

Figures 2 and 3 show the transverse momentum spectra of $J/\psi$ mesons produced in *pp* collisions at $\sqrt{s} = 8$ and 13 TeV, respectively. Figures 2(a) and 2(b) (Figures 3(a) and 3(b)) present the results of the prompt $J/\psi$ and $J/\psi$ from *b*, respectively. The symbols represent the experimental data measured by the LHCb Collaboration [16, 18] at the LHC and the curves are our fitted results. The values of free parameters and $\chi^2/dof$ corresponding to each curve in Figures 2 and 3 are listed in Table 1, which will be discussed later. One can see again that the experimental data measured by the LHCb Collaboration are well fitted by the two-component statistical model.

The transverse momentum spectra of $J/\psi$ mesons produced in $p$-Pb collisions at center-of-mass energy per nucleon pair $\sqrt{s_{NN}} = 5$ TeV are displayed in Figure 4. Figures 4(a) and 4(b) present the results of the prompt $J/\psi$ and $J/\psi$ from $b$, respectively. The symbols represent the experimental data measured by the LHCb Collaboration [17] at the LHC and the curves are our fitted results. The values of free parameters and $\chi^2/dof$ corresponding to each curve in Figure 4 are listed in Table 1, which will be discussed later. Once again, the experimental data measured by the LHCb Collaboration are well fitted by the model.

The transverse momentum spectra of $\Upsilon$ mesons ($\Upsilon(1S)$, $\Upsilon(2S)$ and $\Upsilon(3S)$) produced in $pp$ collisions at $\sqrt{s} = 8$ TeV are shown in Figure 5, where $B^{1S}$ ($B^{2S}$ and $B^{3S}$) on the vertical axis denotes the branch ratio. The symbols represent the experimental data measured by the LHCb Collaboration [16] and the curves are our fitted results. The values of free parameters and $\chi^2/dof$ corresponding to each curve in Figure 5 are listed in Table 1. Once more, the experimental data measured by the LHCb Collaboration are well fitted by the model.

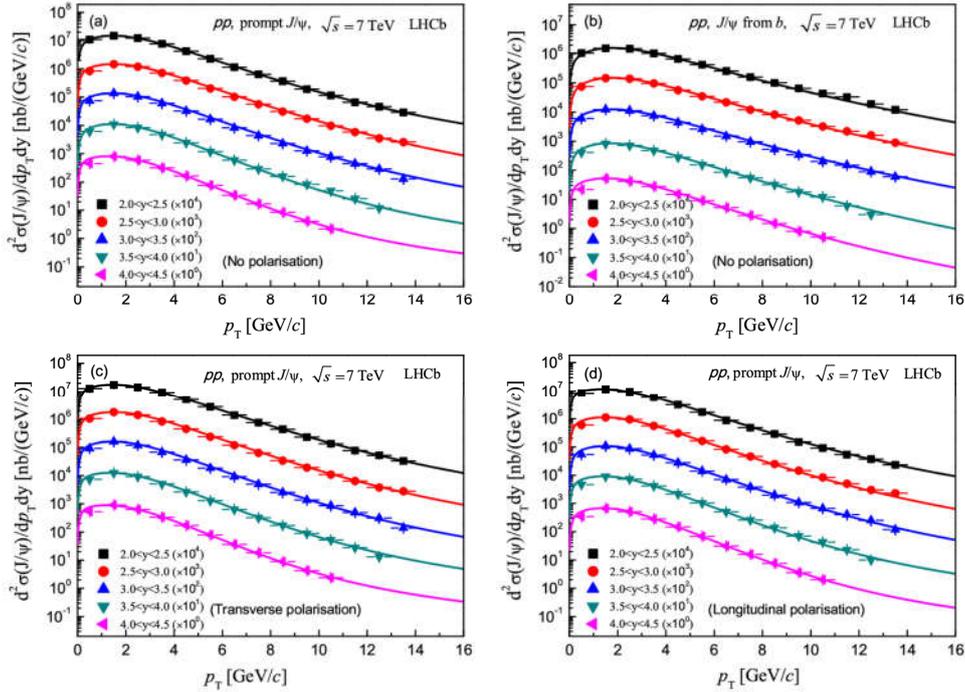

Figure 1: Transverse momentum spectra of (a) prompt $J/\psi$ with no polarisation, (b) $J/\psi$ from $b$ with no polarisation, (c) prompt $J/\psi$ with full transverse polarisation, and (d) prompt $J/\psi$ with full longitudinal polarisation in $pp$ collision at $\sqrt{s} = 7$ TeV. The symbols with the error bars represent the experimental data with the quadratic sums of the statistical and systematic uncertainties measured by the LHCb Collaboration [15] in different rapidity ranges and scaled by different amounts marked in the panels. The curves are our fitted results.

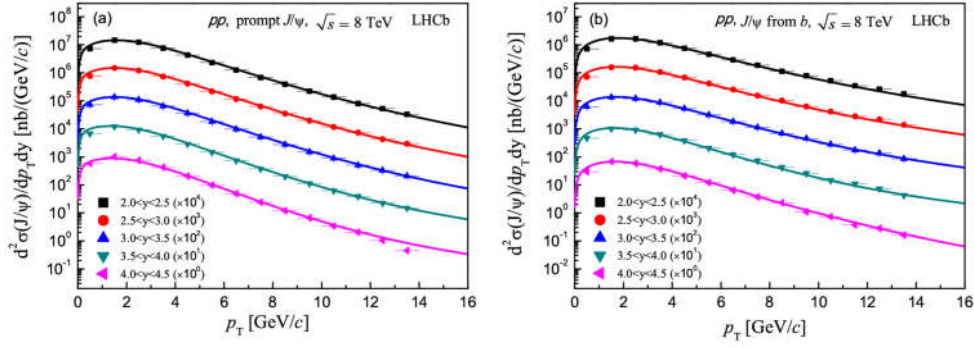

Figure 2: Transverse momentum spectra of (a) prompt $J/\psi$ and (b) $J/\psi$ from $b$ in $pp$ collisions at $\sqrt{s} = 8$ TeV. The symbols with the error bars represent the experimental data with the quadratic sums of the statistical and systematic uncertainties measured by the LHCb Collaboration [16]. The curves are our fitted results.

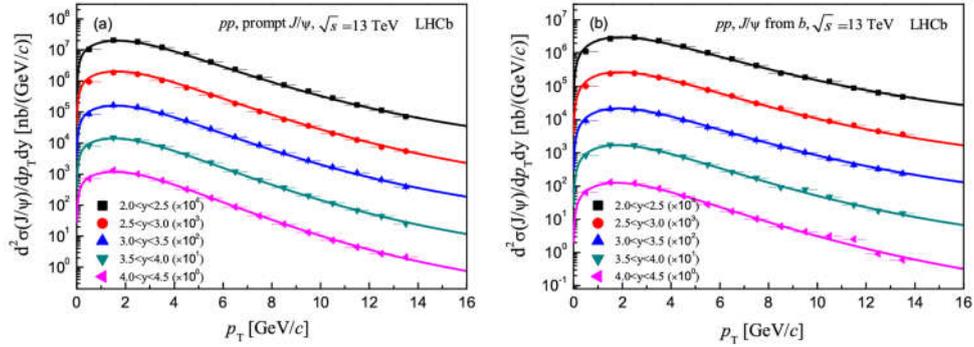

Figure 3: The same as Figure 2, but showing the results of $J/\psi$ mesons produced in $pp$ collisions at $\sqrt{s} = 13$ TeV. The data with the statistical uncertainties are quoted from ref. [18].

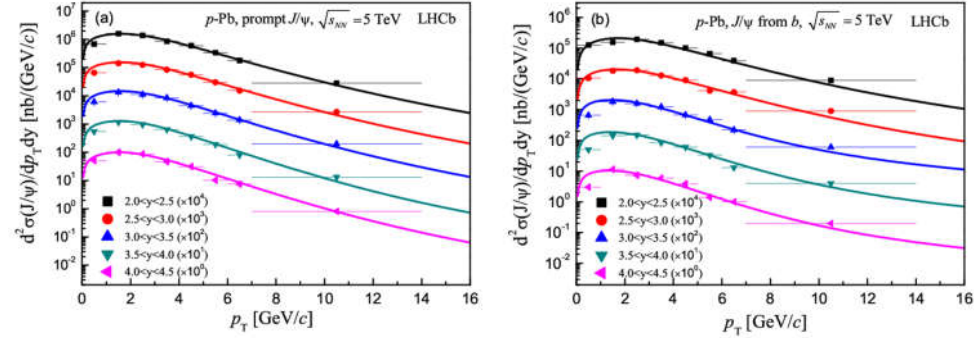

Figure 4: The same as Figure 2, but showing the results of $J/\psi$ mesons in $p$-Pb collisions at $\sqrt{s_{NN}} = 5$ TeV. The data with the quadratic sums of the statistical and systematic uncertainties are quoted from ref. [17].

To see clearly the relationships between the free parameters ($T$, $q$, $p_0$, and $n$) and rapidity, we plot the parameter values listed in Table 1 in Figures 6-9, respectively. In the four figures, the symbols represent the parameters and the lines are our fitted results, though some of them do not obey the linear functions. In Figure 9, some error bars are smaller than the symbol size due to wide coordinate range. The intercepts, slopes, and $\chi^2/dof$ corresponding to the lines in Figures 6-9 are listed in Table 2. One can see that, in the error range, the parameter $T$ does not show an obvious change or has a slight decrease in most cases, the parameters $q$ and $p_0$ appear to decrease, and the parameter *n* does not show an obvious change or has a slight increase, with the increase of rapidity.

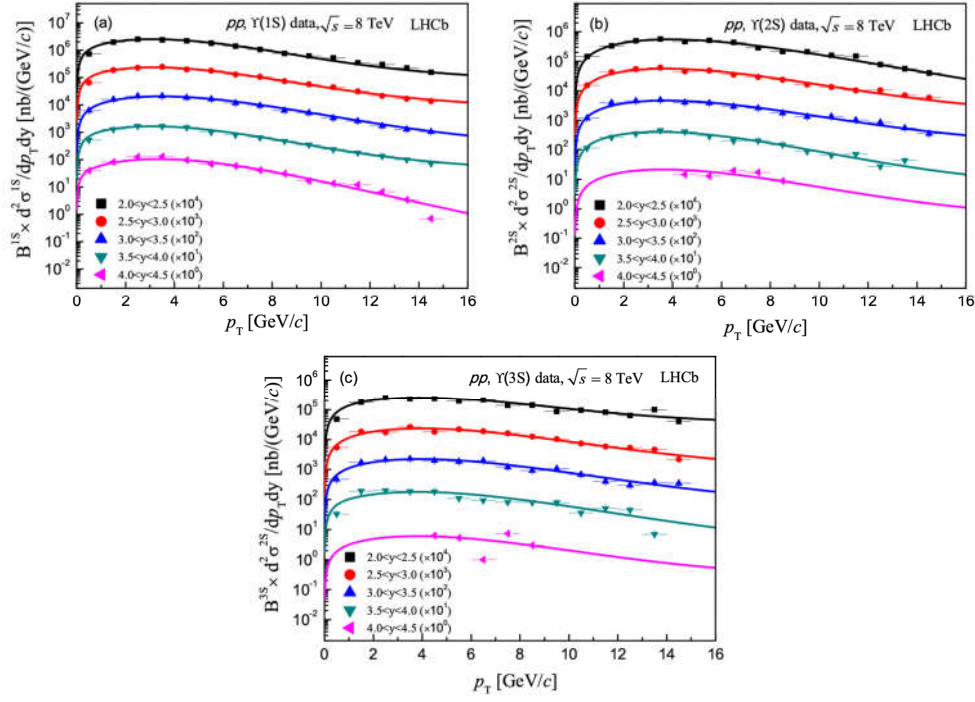

Figure 5: Transverse momentum spectra of $\Upsilon$ mesons ($\Upsilon(1S)$, $\Upsilon(2S)$ and $\Upsilon(3S)$) produced in $pp$ collision at $\sqrt{s} = 8$ TeV. The symbols with the error bars represent the experimental data with the quadratic sums of the statistical and systematic uncertainties measured by the LHCb Collaboration [16]. The curves are our fitted results.

Table 1: Values of parameters and $\chi^2/dof$ corresponding to the curves in Figures 1–5.

| Figure | Type | $k$ | $T$ (GeV) | $q$ | $p_0$ (GeV/$c$) | $n$ | $\chi^2/dof$ |
|---|---|---|---|---|---|---|---|
| Figure 1(a) | 2.0<y<2.5 | 0.890±0.010 | 0.440±0.007 | 1.085±0.006 | 1.622±0.008 | 5.000±0.020 | 0.096 |
| | 2.5<y<3.0 | 0.900±0.005 | 0.435±0.005 | 1.085±0.006 | 1.622±0.008 | 5.050±0.017 | 0.164 |
| | 3.0<y<3.5 | 0.888±0.003 | 0.432±0.008 | 1.080±0.005 | 1.625±0.009 | 5.056±0.024 | 0.614 |
| | 3.5<y<4.0 | 0.895±0.005 | 0.440±0.005 | 1.066±0.007 | 1.623±0.007 | 5.050±0.018 | 0.450 |
| | 4.0<y<4.5 | 0.865±0.007 | 0.438±0.009 | 1.060±0.009 | 1.605±0.013 | 5.031±0.019 | 0.317 |
| Figure 1(b) | 2.0<y<2.5 | 0.902±0.005 | 0.468±0.007 | 1.113±0.007 | 1.627±0.017 | 5.037±0.017 | 0.533 |
| | 2.5<y<3.0 | 0.895±0.003 | 0.462±0.008 | 1.109±0.006 | 1.606±0.009 | 5.050±0.020 | 0.296 |
| | 3.0<y<3.5 | 0.892±0.003 | 0.455±0.007 | 1.108±0.007 | 1.600±0.010 | 5.036±0.023 | 0.414 |
| | 3.5<y<4.0 | 0.893±0.005 | 0.453±0.007 | 1.097±0.008 | 1.603±0.010 | 5.026±0.012 | 0.226 |
| | 4.0<y<4.5 | 0.897±0.006 | 0.440±0.008 | 1.092±0.008 | 1.588±0.018 | 5.030±0.020 | 0.276 |
| Figure 1(c) | 2.0<y<2.5 | 0.900±0.006 | 0.443±0.007 | 1.086±0.008 | 1.607±0.007 | 5.000±0.015 | 0.075 |
| | 2.5<y<3.0 | 0.896±0.003 | 0.440±0.005 | 1.081±0.006 | 1.588±0.012 | 5.000±0.017 | 0.485 |
| | 3.0<y<3.5 | 0.894±0.008 | 0.435±0.008 | 1.076±0.008 | 1.600±0.005 | 5.000±0.020 | 0.392 |
| | 3.5<y<4.0 | 0.892±0.006 | 0.430±0.007 | 1.070±0.005 | 1.603±0.005 | 5.008±0.018 | 0.397 |
| | 4.0<y<4.5 | 0.875±0.008 | 0.430±0.006 | 1.060±0.009 | 1.588±0.010 | 5.003±0.013 | 0.150 |
| Figure 1(d) | 2.0<y<2.5 | 0.896±0.004 | 0.439±0.005 | 1.088±0.007 | 1.602±0.008 | 5.005±0.018 | 0.557 |
| | 2.5<y<3.0 | 0.899±0.008 | 0.420±0.007 | 1.086±0.009 | 1.582±0.012 | 5.010±0.015 | 0.324 |
| | 3.0<y<3.5 | 0.895±0.007 | 0.423±0.007 | 1.082±0.007 | 1.612±0.012 | 5.000±0.015 | 0.438 |
| | 3.5<y<4.0 | 0.893±0.008 | 0.430±0.006 | 1.070±0.008 | 1.608±0.007 | 5.008±0.017 | 0.517 |
| | 4.0<y<4.5 | 0.885±0.005 | 0.426±0.008 | 1.065±0.008 | 1.583±0.010 | 5.012±0.013 | 0.330 |
| Figure 2(a) | 2.0<y<2.5 | 0.896±0.002 | 0.424±0.004 | 1.093±0.005 | 1.605±0.008 | 5.000±0.020 | 0.258 |
| | 2.5<y<3.0 | 0.886±0.004 | 0.420±0.006 | 1.090±0.005 | 1.612±0.005 | 5.003±0.015 | 0.245 |
| | 3.0<y<3.5 | 0.890±0.005 | 0.425±0.005 | 1.085±0.007 | 1.603±0.005 | 5.007±0.013 | 0.179 |
| | 3.5<y<4.0 | 0.886±0.004 | 0.421±0.004 | 1.078±0.007 | 1.600±0.005 | 5.012±0.016 | 0.413 |
| | 4.0<y<4.5 | 0.889±0.003 | 0.430±0.007 | 1.072±0.006 | 1.600±0.007 | 5.022±0.016 | 0.899 |
| Figure 2(b) | 2.0<y<2.5 | 0.870±0.004 | 0.443±0.005 | 1.120±0.005 | 1.627±0.007 | 4.988±0.018 | 2.087 |
| | 2.5<y<3.0 | 0.852±0.003 | 0.441±0.004 | 1.115±0.005 | 1.622±0.005 | 4.992±0.016 | 2.057 |
| | 3.0<y<3.5 | 0.850±0.003 | 0.435±0.005 | 1.109±0.007 | 1.623±0.008 | 4.825±0.015 | 1.321 |
| | 3.5<y<4.0 | 0.802±0.004 | 0.441±0.006 | 1.093±0.007 | 1.613±0.008 | 4.867±0.017 | 2.980 |
| | 4.0<y<4.5 | 0.847±0.003 | 0.440±0.006 | 1.095±0.005 | 1.605±0.010 | 5.000±0.021 | 0.902 |
| Figure 3(a) | 2.0<y<2.5 | 0.860±0.004 | 0.445±0.005 | 1.094±0.006 | 1.626±0.006 | 4.095±0.023 | 1.221 |
| | 2.5<y<3.0 | 0.867±0.003 | 0.452±0.006 | 1.089±0.005 | 1.650±0.008 | 4.452±0.018 | 1.569 |
| | 3.0<y<3.5 | 0.868±0.005 | 0.442±0.007 | 1.088±0.007 | 1.635±0.007 | 4.380±0.020 | 1.937 |
| | 3.5<y<4.0 | 0.865±0.003 | 0.435±0.005 | 1.085±0.007 | 1.630±0.008 | 4.630±0.024 | 1.070 |
| | 4.0<y<4.5 | 0.858±0.003 | 0.430±0.007 | 1.082±0.008 | 1.608±0.010 | 4.672±0.026 | 1.062 |
| Figure 3(b) | 2.0<y<2.5 | 0.886±0.004 | 0.522±0.005 | 1.116±0.006 | 1.635±0.005 | 3.150±0.015 | 0.061 |
| | 2.5<y<3.0 | 0.867±0.005 | 0.512±0.007 | 1.109±0.006 | 1.653±0.008 | 3.352±0.017 | 0.886 |
| | 3.0<y<3.5 | 0.887±0.003 | 0.512±0.006 | 1.108±0.007 | 1.644±0.007 | 3.365±0.017 | 0.308 |
| | 3.5<y<4.0 | 0.865±0.005 | 0.510±0.005 | 1.100±0.007 | 1.641±0.007 | 3.640±0.020 | 0.464 |
| | 4.0<y<4.5 | 0.885±0.006 | 0.500±0.007 | 1.098±0.007 | 1.628±0.008 | 3.972±0.020 | 0.638 |
| Figure 4(a) | 2.0<y<2.5 | 0.880±0.003 | 0.432±0.003 | 1.106±0.006 | 1.607±0.005 | 5.012±0.023 | 1.372 |
| | 2.5<y<3.0 | 0.878±0.004 | 0.434±0.004 | 1.102±0.007 | 1.608±0.007 | 4.998±0.025 | 1.788 |
| | 3.0<y<3.5 | 0.883±0.003 | 0.430±0.004 | 1.095±0.009 | 1.610±0.005 | 4.995±0.025 | 1.816 |
| | 3.5<y<4.0 | 0.884±0.005 | 0.432±0.004 | 1.085±0.007 | 1.608±0.006 | 4.990±0.020 | 2.044 |
| | 4.0<y<4.5 | 0.882±0.005 | 0.430±0.005 | 1.088±0.007 | 1.607±0.008 | 5.010±0.027 | 1.639 |
| Figure 4(b) | 2.0<y<2.5 | 0.858±0.003 | 0.428±0.004 | 1.126±0.008 | 1.623±0.008 | 3.956±0.024 | 0.715 |
| | 2.5<y<3.0 | 0.839±0.003 | 0.430±0.002 | 1.122±0.008 | 1.608±0.007 | 3.998±0.027 | 0.739 |
| | 3.0<y<3.5 | 0.705±0.004 | 0.425±0.004 | 1.095±0.007 | 1.601±0.005 | 3.977±0.023 | 3.504 |
| | 3.5<y<4.0 | 0.717±0.003 | 0.432±0.003 | 1.092±0.006 | 1.603±0.007 | 3.890±0.030 | 6.600 |
| | 4.0<y<4.5 | 0.715±0.005 | 0.427±0.003 | 1.088±0.007 | 1.603±0.006 | 4.010±0.030 | 6.856 |
| Figure 5(a) | 2.0<y<2.5 | 0.776±0.002 | 0.537±0.003 | 1.054±0.006 | 1.556±0.013 | 2.355±0.010 | 1.325 |
| | 2.5<y<3.0 | 0.780±0.003 | 0.530±0.005 | 1.056±0.004 | 1.568±0.018 | 2.338±0.012 | 0.490 |
| | 3.0<y<3.5 | 0.865±0.005 | 0.545±0.005 | 1.065±0.007 | 1.541±0.008 | 2.372±0.012 | 0.282 |
| | 3.5<y<4.0 | 0.776±0.003 | 0.539±0.004 | 1.052±0.004 | 1.563±0.012 | 2.490±0.015 | 0.503 |
| | 4.0<y<4.5 | 0.903±0.005 | 0.520±0.005 | 1.058±0.004 | 1.433±0.023 | 4.052±0.022 | 3.266 |
| Figure 5(b) | 2.0<y<2.5 | 0.912±0.004 | 0.567±0.005 | 1.084±0.006 | 1.756±0.009 | 2.655±0.015 | 0.855 |
| | 2.5<y<3.0 | 0.890±0.003 | 0.530±0.005 | 1.079±0.007 | 1.791±0.012 | 2.000±0.015 | 0.921 |
| | 3.0<y<3.5 | 0.870±0.003 | 0.568±0.006 | 1.075±0.008 | 1.815±0.018 | 2.000±0.013 | 0.976 |
| | 3.5<y<4.0 | 0.907±0.005 | 0.533±0.007 | 1.066±0.008 | 1.708±0.025 | 2.430±0.017 | 1.801 |
| | 4.0<y<4.5 | 0.896±0.003 | 0.578±0.005 | 1.068±0.008 | 1.733±0.010 | 2.052±0.010 | 1.219 |
| Figure 5(c) | 2.0<y<2.5 | 0.705±0.003 | 0.550±0.007 | 1.074±0.005 | 1.726±0.018 | 1.705±0.017 | 2.793 |
| | 2.5<y<3.0 | 0.716±0.002 | 0.552±0.008 | 1.079±0.007 | 1.788±0.024 | 2.000±0.010 | 1.146 |
| | 3.0<y<3.5 | 0.725±0.005 | 0.562±0.008 | 1.078±0.007 | 1.755±0.015 | 2.185±0.015 | 2.025 |
| | 3.5<y<4.0 | 0.750±0.003 | 0.573±0.005 | 1.076±0.006 | 1.738±0.018 | 2.630±0.020 | 7.854 |
| | 4.0<y<4.5 | 0.736±0.005 | 0.558±0.006 | 1.072±0.008 | 1.710±0.025 | 2.052±0.012 | 6.918 |

It should be noted that Figure 6(d) shows a slight increase of $T$ for $\Upsilon(2S)$ and $\Upsilon(3S)$, and a slight decrease of $T$ for $\Upsilon(1S)$, with the increase of rapidity. An average weighted by different yields will result in a slight decrease of $T$ with the increase of rapidity. The rapidity dependent $p_0$ and $n$ confirm the prediction of pQCD which gives the inverse power-law spectra being rapidity dependent [32-34].

The meanings of parameters can be explained by us. The invariant or slight decreasing temperature parameter renders that the excitation degree of the interacting system keeps invariant or slight decreasing trend with the increase of rapidity. The temperature is not the "real" temperature at the stage of kinetic freeze-out, but the effective temperature in which the contribution of flow effect is not excluded. Even the flow effect is excluding, the kinetic freeze-out temperature from the spectra of heavy quarkonia is much higher than that from the spectra of light particles. This means that the heavy quarkonia produce much earlier than light particles in the collision process.

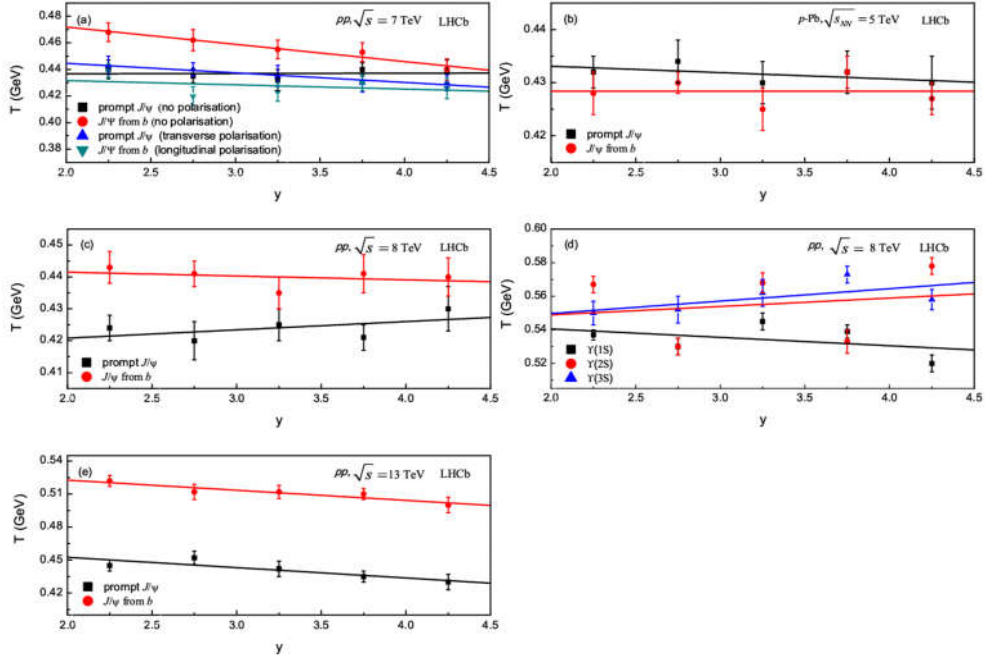

Figure 6: The relationship between *T* and *y* for $J/\psi$ and $\Upsilon$ mesons produced in *pp* and *p*-Pb collisions at the LHC. The symbols are quoted in Tables 1, and the lines are our fitted results.

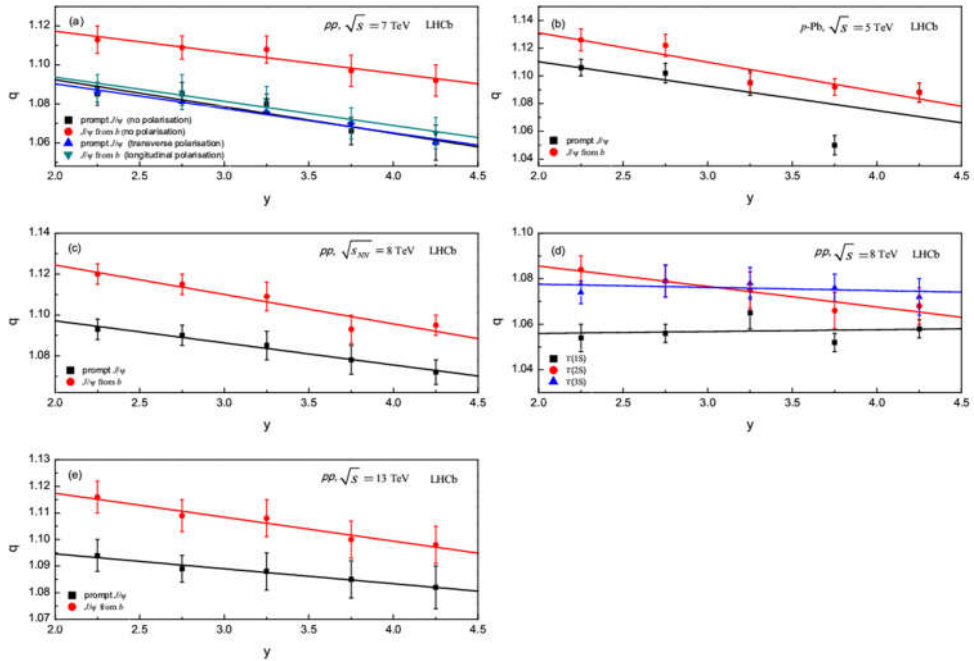

Figure 7: The same as Figure 6, but showing the relationship between *q* and *y*.

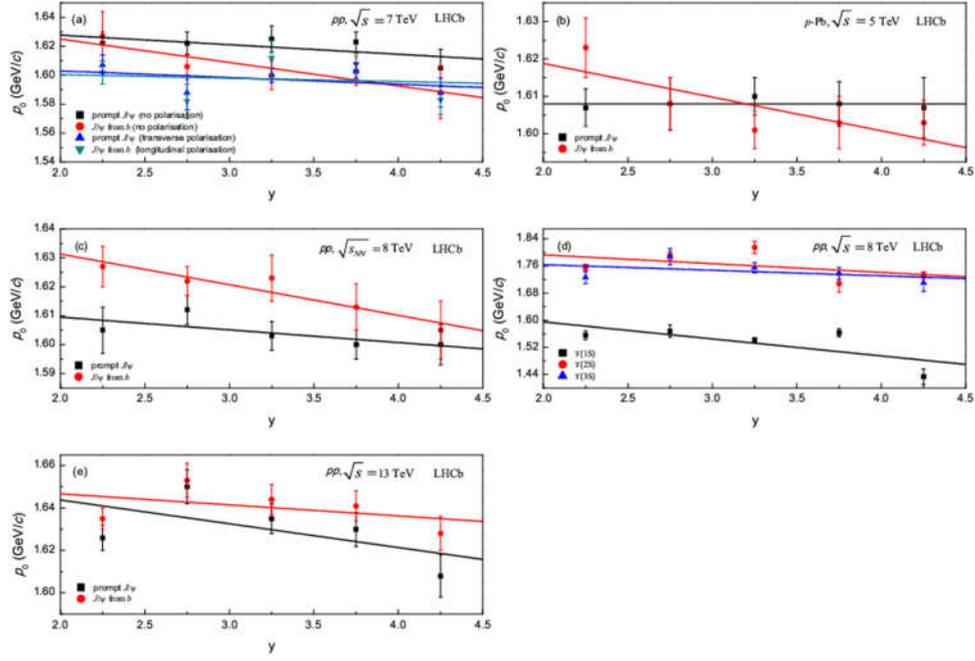

Figure 8: The same as Figure 6, but showing the relationship between $p_0$ and $y$.

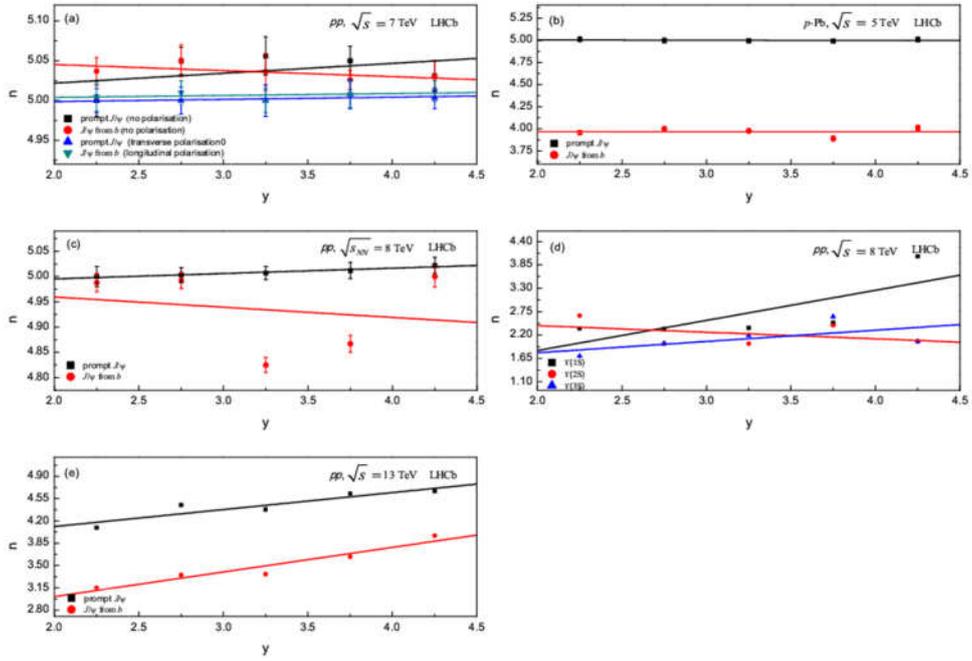

Figure 9: The same as Figure 6, but showing the relationship between n and y.

Table 2: Values of intercepts, slopes, and $\chi^2/dof$ corresponding to the lines in Figures 6-9.

| Figure | Type | Intercept | Slope | $\chi^2/dof$ |
|---|---|---|---|---|
| Figure 6(a) | prompt $J/\psi$ (no polarisation) | $0.436\pm0.007$ | $0.000\pm0.002$ | 0.363 |
| | $J/\psi$ from $b$ (no polarisation) | $0.498\pm0.005$ | $-0.013\pm0.002$ | 0.144 |
| | prompt $J/\psi$ (transverse polarisation) | $0.459\pm0.003$ | $-0.007\pm0.001$ | 0.062 |
| | $J/\psi$ from $b$ (longitudinal polarisation) | $0.438\pm0.015$ | $-0.003\pm0.004$ | 1.778 |
| Figure 6(b) | prompt $J/\psi$ | $0.436\pm0.003$ | $-0.001\pm0.001$ | 0.168 |
| | $J/\psi$ from b | $0.428\pm0.006$ | $0.000\pm0.002$ | 1.010 |
| Figure 6(c) | prompt $J/\psi$ | $0.416\pm0.007$ | $0.003\pm0.002$ | 0.686 |
| | $J/\psi$ from b | $0.444\pm0.006$ | $-0.001\pm0.002$ | 0.417 |
| Figure 6(d) | $\Upsilon(1S)$ | $0.550\pm0.018$ | $-0.005\pm0.006$ | 4.572 |
| | $\Upsilon(2S)$ | $0.539\pm0.046$ | $0.005\pm0.014$ | 20.526 |
| | $\Upsilon(3S)$ | $0.535\pm0.015$ | $0.007\pm0.004$ | 2.189 |
| Figure 6(e) | prompt $J/\psi$ | $0.471\pm0.009$ | $-0.009\pm0.003$ | 0.791 |
| | $J/\psi$ from $b$ | $0.541\pm0.006$ | $-0.009\pm0.002$ | 0.318 |
| Figure 7(a) | prompt $J/\psi$ (no polarisation) | $1.120\pm0.008$ | $-0.014\pm0.002$ | 0.338 |
| | $J/\psi$ from $b$ (no polarisation) | $1.139\pm0.005$ | $-0.011\pm0.002$ | 0.153 |
| | prompt $J/\psi$ (transverse polarisation) | $1.116\pm0.003$ | $-0.013\pm0.001$ | 0.073 |
| | $J/\psi$ from $b$ (longitudinal polarisation) | $1.118\pm0.006$ | $-0.012\pm0.002$ | 0.179 |
| Figure 7(b) | prompt $J/\psi$ | $1.145\pm0.037$ | $-0.018\pm0.011$ | 8.300 |
| | $J/\psi$ from b | $1.174\pm0.014$ | $-0.021\pm0.004$ | 1.049 |
| Figure 7(c) | prompt $J/\psi$ | $1.119\pm0.002$ | $-0.011\pm0.001$ | 0.059 |
| | $J/\psi$ from b | $1.153\pm0.008$ | $-0.014\pm0.002$ | 0.462 |
| Figure 7(d) | $\Upsilon(1S)$ | $1.054\pm0.010$ | $0.001\pm0.003$ | 1.096 |
| | $\Upsilon(2S)$ | $1.104\pm0.005$ | $-0.009\pm0.002$ | 0.120 |
| | $\Upsilon(3S)$ | $1.080\pm0.006$ | $-0.001\pm0.002$ | 0.250 |
| Figure 7(e) | prompt $J/\psi$ | $1.106\pm0.002$ | $-0.006\pm0.001$ | 0.033 |
| | $J/\psi$ from b | $1.135\pm0.003$ | $-0.009\pm0.001$ | 0.079 |
| Figure 8(a) | prompt $J/\psi$ (no polarisation) | $1.641\pm0.013$ | $-0.007\pm0.004$ | 0.659 |
| | $J/\psi$ from $b$ (no polarisation) | $1.657\pm0.013$ | $-0.016\pm0.004$ | 0.447 |
| | prompt $J/\psi$ (transverse polarisation) | $1.612\pm0.017$ | $-0.005\pm0.005$ | 1.540 |
| | $J/\psi$ from $b$ (longitudinal polarisation) | $1.605\pm0.029$ | $-0.002\pm0.009$ | 2.584 |
| Figure 8(b) | prompt $J/\psi$ | $1.608\pm0.003$ | $0.000\pm0.001$ | 0.072 |
| | $J/\psi$ from b | $1.637\pm0.012$ | $-0.009\pm0.004$ | 1.088 |
| Figure 8(c) | prompt $J/\psi$ | $1.618\pm0.007$ | $-0.004\pm0.002$ | 0.566 |
| | $J/\psi$ from b | $1.652\pm0.006$ | $-0.011\pm0.002$ | 0.190 |
| Figure 8(d) | $\Upsilon(1S)$ | $1.695\pm0.084$ | $-0.050\pm0.025$ | 10.642 |
| | $\Upsilon(2S)$ | $1.844\pm0.080$ | $-0.026\pm0.024$ | 8.408 |
| | $\Upsilon(3S)$ | $1.797\pm0.057$ | $-0.016\pm0.017$ | 2.304 |
| Figure 8(e) | prompt $J/\psi$ | $1.666\pm0.026$ | $-0.011\pm0.008$ | 3.927 |
| | $J/\psi$ from b | $1.657\pm0.018$ | $-0.005\pm0.005$ | 2.416 |
| Figure 9(a) | prompt $J/\psi$ (no polarisation) | $4.997\pm0.044$ | $0.012\pm0.013$ | 1.497 |
| | $J/\psi$ from $b$ (no polarisation) | $5.060\pm0.014$ | $-0.008\pm0.004$ | 0.223 |
| | prompt $J/\psi$ (transverse polarisation) | $4.993\pm0.006$ | $0.003\pm0.002$ | 0.033 |
| | $J/\psi$ from $b$ (longitudinal polarisation) | $4.999\pm0.009$ | $0.002\pm0.003$ | 0.112 |
| Figure 9(b) | prompt $J/\psi$ | $5.009\pm0.020$ | $-0.002\pm0.006$ | 0.215 |
| | $J/\psi$ from b | $3.966\pm0.099$ | $0.000\pm0.030$ | 3.457 |
| Figure 9(c) | prompt $J/\psi$ | $4.974\pm0.004$ | $0.011\pm0.001$ | 0.024 |
| | $J/\psi$ from b | $5.000\pm0.170$ | $-0.020\pm0.051$ | 31.168 |
| Figure 9(d) | $\Upsilon(1S)$ | $0.416\pm1.036$ | $0.709\pm0.311$ | 1450.893 |
| | $\Upsilon(2S)$ | $2.732\pm0.574$ | $-0.155\pm0.173$ | 441.653 |
| | $\Upsilon(3S)$ | $1.254\pm0.557$ | $0.265\pm0.167$ | 402.772 |
| Figure 9(e) | prompt $J/\psi$ | $3.580\pm0.197$ | $0.266\pm0.059$ | 30.391 |
| | $J/\psi$ from $b$ | $2.240\pm0.187$ | $0.386\pm0.056$ | 33.711 |
| Figure 10(a) | prompt $J/\psi$ (no polarisation) | $2.960\pm0.110$ | $-0.161\pm0.033$ | 0.144 |
|  | $J/\psi$ from $b$ (no polarisation) | 3.692±0.098 | -0.231±0.030 | 2.259 |
|  | prompt $J/\psi$ (transverse polarisation) | 3.038±0.071 | -0.191±0.021 | 0.079 |
|  | $J/\psi$ from $b$ (longitudinal polarisation) | 2.855±0.086 | -0.127±0.026 | 0.076 |
| Figure 10(b) | prompt $J/\psi$ | 3.273±0.087 | -0.170±0.026 | 3.307 |
|  | $J/\psi$ from b | 3.867±0.190 | -0.290±0.057 | 18.317 |
| Figure 10(c) | prompt $J/\psi$ | 3.031±0.060 | -0.157±0.018 | 0.031 |
|  | $J/\psi$ from b | 3.669±0.099 | -0.221±0.030 | 4.616 |
| Figure 10(d) | $\Upsilon(1S)$ | 5.429±0.278 | -0.155±0.083 | 0.029 |
|  | $\Upsilon(2S)$ | 5.529±0.697 | 0.086±0.021 | 0.025 |
|  | $\Upsilon(3S)$ | 6.730±0.254 | -0.252±0.076 | 0.015 |
| Figure 10(e) | prompt $J/\psi$ | 3.126±0.026 | -0.134±0.008 | 0.002 |
|  | $J/\psi$ from $b$ | 3.995±0.070 | -0.217±0.021 | 0.048 |
| Figure 11(a) | prompt $J/\psi$ (no polarisation) | 2.597±0.082 | -0.151±0.025 | 0.020 |
|  | $J/\psi$ from $b$ (no polarisation) | 3.198±0.078 | -0.200±0.024 | 0.076 |
|  | prompt $J/\psi$ (transverse polarisation) | 2.551±0.178 | -0.129±0.054 | 0.201 |
|  | $J/\psi$ from $b$ (longitudinal polarisation) | 2.543±0.059 | -0.131±0.018 | 0.011 |
| Figure 11(b) | prompt $J/\psi$ | 2.847±0.080 | -0.151±0.024 | 0.149 |
|  | $J/\psi$ from b | 3.379±0.163 | -0.249±0.049 | 5.393 |
| Figure 11(c) | prompt $J/\psi$ | 2.620±0.040 | -0.135±0.012 | 0.005 |
|  | $J/\psi$ from b | 3.215±0.094 | -0.204±0.028 | 0.195 |
| Figure 11(d) | $\Upsilon(1S)$ | 4.537±0.238 | -0.134±0.072 | 0.039 |
|  | $\Upsilon(2S)$ | 5.039±0.273 | -0.115±0.082 | 0.015 |
|  | $\Upsilon(3S)$ | 5.527±0.223 | -0.205±0.067 | 0.023 |
| Figure 11(e) | prompt $J/\psi$ | 2.574±0.264 | -0.057±0.079 | 0.138 |
|  | $J/\psi$ from $b$ | 3.513±0.064 | -0.201±0.019 | 0.014 |

All values of the entropy index are close to 1, which means that the interacting system stays approximately at the (local) equilibrium state, even if in small collision systems such as $pp$ and $p$-Pb collisions. The decreasing entropy index renders that the interacting system reaches to a more equilibrium state in the very forward rapidity region. We believe that the interacting system stays at the (local) equilibrium state in large collision systems such as lead-lead and other nucleus-nucleus collisions.

The decreasing $p_0$ and increasing $n$ renders a narrow $p_T$ range. The present work shows a slightly narrow $p_T$ range in the very forward rapidity region. This means that the scattering strength of the interacting system decreases slightly with the increase of rapidity. This observation confirms the result from the temperature parameter.

The contribution ratios of the first component (the Tsallis statistics) are in the range from 0.705 to 0.912, which are listed only in Table 1 and not shown in plot to avoid trivialness. The main contribution ratios reflect the strong power of the Tsallis statistics in the fitting process for the $p_T$ spectra of heavy quarkonia in various rapidity regions. Meanwhile, the contribution ratios $(1-k)$ of the second component (the Hagedorn function) are considerable. The contribution ratios do not show particular behaviors, but almost invariant with rapidity. This means that the impact between the two "participant" quarks is very violent. The effects of other factors such as the rapidity region are not dominant.

We would like to point out that the Hagedorn function is indeed needed, though the Tsallis statistics has power law tail in high $p_T$ region. In fact, if we use the Tsallis statistics to fit the spectra in high $p_T$ region, the fit in low $p_T$ region will be failed. Contrarily, if the low $p_T$ region is fixed, the high $p_T$ region will be on the wrong way. In our opinion, the Tsallis statistics should fit the spectra from 0 to the range as widely as possible. The Hagedorn function should fit the spectra in high $p_T$ region, though its contribution is from 0 to high $p_T$ region.

Comparing with that in *pp* collisions, the parameters from the spectra in *p*-Pb collisions do not show particular behaviors. This means that the cold nuclear effect affects mainly the normalizations of $p_T$ spectra of heavy quarkonia, but not the shapes. In fact, the heavy quarkonia are produced in the process of violent impact between two "participant" quarks in the considered collisions. Not only the spectator nucleons but also the "spectator" quarks do not affect largely the shapes of $p_T$ spectra of heavy quarkonia. Naturally, the cold nuclear effect in lead-lead and other nucleus-nucleus collisions has no large effect on the shapes of $p_T$ spectra of heavy quarkonia.

It should be noted that a given free parameter for prompt $J/\psi$ and $J/\psi$ from *b* leads to similar result or small difference at one energy, and large difference at another energy. If small difference is explained by statistical fluctuation in the data, large difference can be explained by dynamical reason. For example, the results for prompt $J/\psi$ and $J/\psi$ from *b* at 7 and 8 TeV lead to similar $p_0$ and *n*, and the results at 13 TeV lead to different *n* in both cases. It is possible that the dynamical mechanism at 13 TeV is different from that at 7 and 8 TeV due to different energies. The large difference should be studied in the near future by the more accurate pQCD method.

In addition, in which concerns prompt $J/\psi$, most of free parameters show no or very small dependence on rapidity, except maybe at 13 TeV. This seems that the difference, if existent, shown by the data is contained either in the normalization or the previously mentioned subtracted mid-value of the given rapidity range. In fact, both the factors do not affect the free parameters which are only determined by the shapes of $p_T$ spectra. The treatment shifted the given rapidity range to the mid-rapidity region is necessary due to the fact that the kinetic energy of directional movement should not be included in the temperature.

To analyze further the behaviors of parameters, Figures 10 and 11 present the dependences of mean $p_T$ ($\langle p_T \rangle$) and ratio of root-mean-square $p_T$ ($\sqrt{\langle p_T^2 \rangle}$) to $\sqrt{2}$ on rapidity respectively. The symbols represent the values of $\langle p_T \rangle$ and $\sqrt{\langle p_T^2 \rangle/2}$ obtained from the curves in Figures 1-5. The lines are the fitted results of linear functions, though some of them do not obey the linear relationship. The intercepts, slopes, and $\chi^2/dof$ corresponding to these lines are listed in Table 2. One can see the decreasing trends of the considered quantities with the increase of rapidity. These trends also render decreasing excitation degree of the interacting system with the increase of rapidity.

According to refs. [37-39], if the initial temperature ($T_i$) of the interacting system is approximately described by $\sqrt{\langle p_T^2 \rangle/2}$, Figure 11 shows that $T_i$ decreases when the rapidity increases due to less energy deposition in very forward rapidity region. At the same or similar LHC energy, $T_i$ extracted from the spectra of $J/\psi$ (or $\Upsilon$) mesons is about 6 (or 12) times of that (~0.4 GeV) extracted from the spectra of pion mesons [40]. As the quantities which are independent of models, both $\langle p_T \rangle$ and $T_i$ are very important in the understanding the excitation degree of interacting system. More investigations on $\langle p_T \rangle$ and $T_i$ are needed due to their importance.

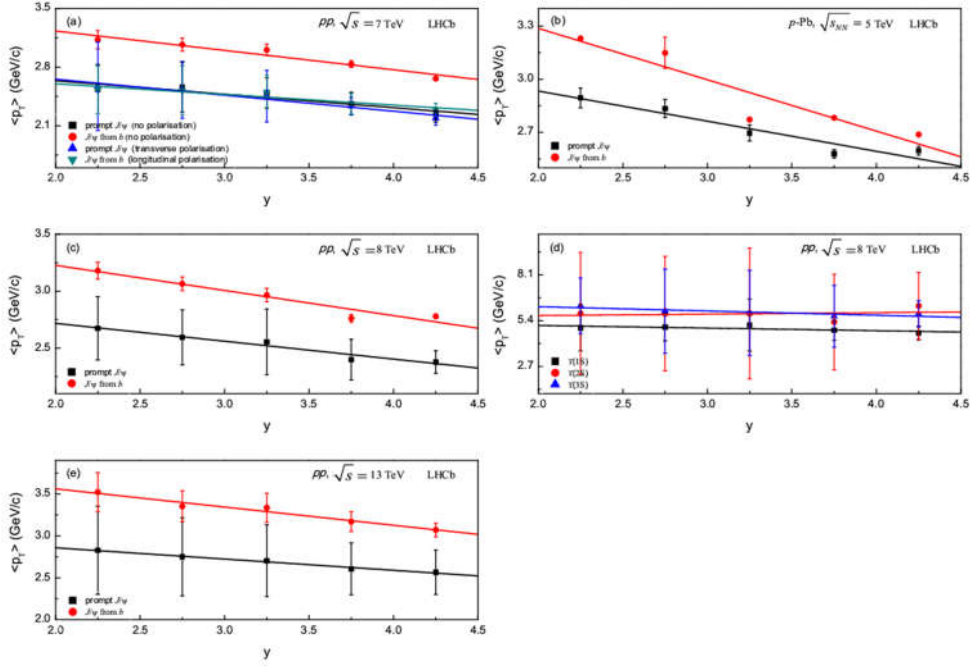

Figure 10: The same as Figure 6, but showing the relationship between $\langle p_T \rangle$ and $y$.

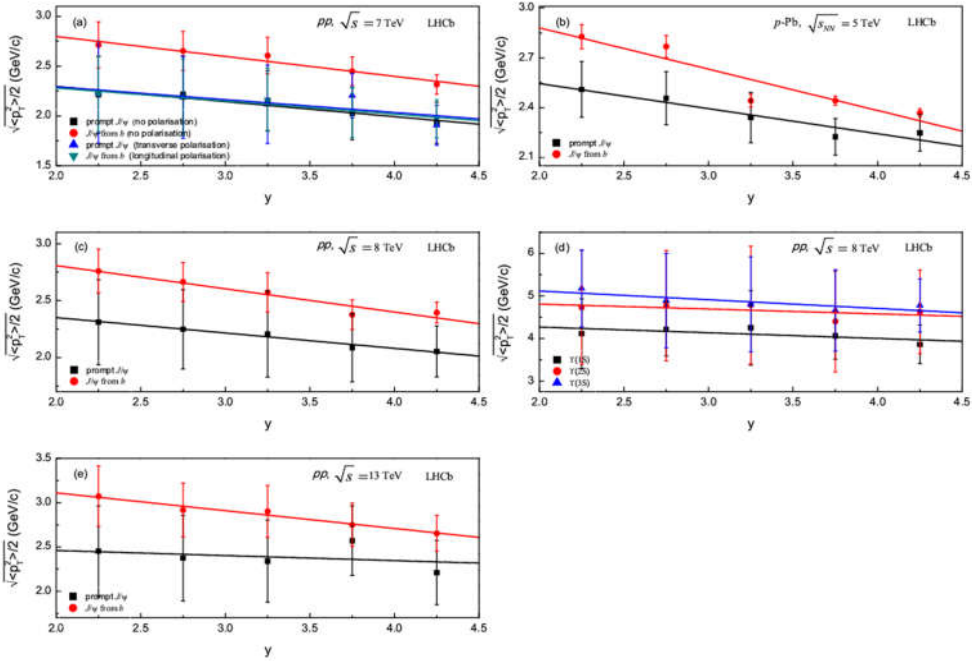

Figure 11: The same as Figure 6, but showing the relationship between $\sqrt{\langle p_T^2 \rangle/2}$ and $y$.

Similar to most of free parameters, the derived quantities $\langle p_T \rangle$ and $T_i$ from the model also show no or very small dependence on rapidity. Although $\langle p_T \rangle$ and $T_i$ are calculated from the model, they depend mainly on the data. In fact, $\langle p_T \rangle$ and $T_i$ are independent of models. The similar dependence on rapidity for most of free parameters and derived quantities renders that the treatment previously mentioned subtracted mid-value of the given rapidity range is correct.

Before conclusions, we would like to point out that the present work is a new analysis on the LHCb data [15-18] analyzed in our recent work [41] which does not fit very well the spectra in high $p_T$ region. In the present work, to fit the spectra in high $p_T$ region, the inverse power-law [32-34] (Hagedorn function [35]) is used as the second component to structure a superposition with the Tsallis statistics [27-31]. Intuitively, the fitted results of the two-component model are obviously improved, though three more parameters are introduced. Anyhow, the behaviors of more parameters are revealed in the present work.

## 4. Conclusions

We summarize here our main observations and conclusions.

The rapidity dependent transverse momentum spectra of heavy quarkonia ($J/\psi$ and $\Upsilon$ mesons) produced in small collision systems (*pp* and *p*-Pb collisions) at high energy ($\sqrt{s}$ ($\sqrt{s_{NN}}$) = 5-13 TeV) have been analyzed by a two-component statistical model which is based on the Tsallis statistics and inverse power-law. The experimental data measured by the LHCb Collaboration at the LHC are well fitted by the model results. The related parameters are obtained and the dependences of parameters on rapidity are analyzed.

The invariant or slight decreasing temperature parameter renders that the excitation degree of the interacting system keeps invariant or slight decreasing trend with the increase of rapidity. The heavy quarkonia produce much earlier than light particles due to very high temperature from the spectra of heavy quarkonia. The considered interacting system stays approximately at the (local) equilibrium state due to the entropy index being close to 1. The decreasing entropy index renders that the system stays at a more equilibrium state in the very forward rapidity region.

A slightly narrow $p_T$ range in the very forward rapidity region is observed due to the decreasing $p_0$ and increasing *n*. This means that the scattering strength of the interacting system decreases slightly with the increase of rapidity. The contribution ratio of the Tsallis statistics is close to 1. This reflects the strong power of the Tsallis statistics in the fitting process for the $p_T$ spectra of heavy quarkonia, in various rapidity regions. The impact between the two "participant" quarks is very violent. Other factors do not play dominant functions.

The cold nuclear effect does not affect largely the production of heavy quarkonia due to the fact that the parameters from the spectra in *p*-Pb collisions do not show particular behaviors, comparing with that in *pp* collisions. Not only the spectator nucleons but also the "spectator" quarks do not affect largely the production of heavy quarkonia. The heavy quarkonia are only produced in the process of violent impact between two "participant" quarks in the considered collisions.

The mean transverse momentum $\langle p_T \rangle$, root-mean-square transverse momentum $\sqrt{\langle p_T^2 \rangle}$, and initial temperature $\sqrt{\langle p_T^2 \rangle/2}$ decreases with the increase of rapidity due to the fact that less energy deposition appears in very forward rapidity region. At the same or similar LHC energy, the initial temperature extracted from the spectra of $J/\psi$ (or $\Upsilon$) mesons is about 6 (or 12) times of that (~0.4 GeV) extracted from the spectra of pion mesons.

## Data Availability

The data used to support the findings of this study are quoted from the mentioned references. As a phenomenological work, this paper does not report new data.

**Compliance with Ethical Standards**

The authors declare that they are in compliance with ethical standards regarding the content of this paper.

**Conflicts of Interest**

The authors declare that they have no conflict of interest regarding the publication of this paper.

**Acknowledgments**

Author L.N.G. acknowledges the financial supports from the National Natural Science Foundation of China under Grant No. 11847003, the Doctoral Scientific Research Foundation of Taiyuan Normal University under Grant No. I170167, and the Doctoral Scientific Research Foundation of Shanxi Province under Grant No. I170269. Other authors acknowledge the financial supports from the National Natural Science Foundation of China under Grant Nos. 11575103 and 11847311, the Shanxi Provincial Natural Science Foundation under Grant No. 201701D121005, and the Fund for Shanxi "1331 Project" Key Subjects Construction.